\documentclass[psfig,a4paper]{article}

\usepackage{amsmath,latexsym}

\newtheorem{definition}{Definition}

\newtheorem{property}{Property}
\def\proof{{\bf{Proof }}} 
\def\fin{\hspace{\stretch{1}}$\Box$}
\def\st{\,:\,}
\newcommand{\eq}[1]{\begin{equation*} #1 \end{equation*}}
\newcommand{\eqa}[1]{\begin{eqnarray*} #1 \end{eqnarray*}}

\def\Tr{{\rm Tr}}
\def\Id{\textbf{1}}
\def\Nat{{\bf N}}
\def\Real{{\bf R}}

\def\N{\{1\ldots n\}}
\def\bin{\{0,1\}}
\def\bases{\{+,\times\}}
\def\bell{\{0,1,2,3\}}
\def\w{w}
\def\kerK{{\mathcal{K}}}
\def\genK{{\mathcal{G}}}
\def\suppl{{\mathcal{S}}}
\def\scal{\cdot}
\def\ortho{\bot}

\newcommand{\rv}[1]{\boldsymbol{#1}}
\newcommand{\pr}[1]{{\rm P}_{\rv{#1}}}
\newcommand{\pc}[2]{{\rm P}_{\rv{#1}\,|\,#2}}

\newcommand{\ket}[2]{ | \, {#1} \rangle_{#2}}
\newcommand{\bra}[2]{\,_{#2} \langle {#1} \,  |}
\newcommand{\proj}[2]{\ket{#1}{#2}\bra{#1}{#2}}

\newcommand{\ketb}[1]{ | \, {#1} \rangle}
\newcommand{\brab}[1]{ \langle {#1} \,  |}
\newcommand{\projb}[1]{\ketb{#1}\brab{#1}}
\newcommand{\scalarb}[2]{ \langle {#1} \,  | \, {#2}  \rangle}

\def\phip{\Phi^+}

\def\a{\vec a}
\def\b{\vec b}
\def\aa{\vec \alpha}
\def\bb{\vec \beta}
\def\e{\vec e}
\def\c{\vec c}
\def\cc{\vec \gamma}
\def\k{\vec\kappa}
\def\d{\vec d}
\def\ttheta{\vec\theta_{\k}}
\def\dw{d_K}
\def\ens{\scriptscriptstyle}
\def\aaE{\alpha_{\ens E}}
\def\aE{a_{\ens E}}
\def\cE{c_{\ens E}}
\def\aaR{\alpha_{\ens R}}
\def\aR{a_{\ens R}}

\def\cR{c_{\ens R}}
\def\aaT{\alpha_{\ens T}}
\def\aT{a_{\ens T}}
\def\cT{c_{\ens T}}
\def\aaS{\alpha_{\ens S}}
\def\bbS{\beta_{\ens S}}
\def\cS{c_{\ens S}}

\def\coR{c_{\ens \overline{R}}}
\def\ScoR{\begin{array}{l}\scriptstyle \coR\st\\ \scriptstyle \cE\in Y_{\aE},\\ \scriptstyle \cT\in X_{\aT}\end{array}}

\def\P{{\mathcal{P}}}
\def\V{{\mathcal{V}}}
\def\C{{\mathcal{C}}}

\def\X{{\mathcal{X}}}
\def\Y{{\mathcal{Y}}}

\def\phiv{\chi_{v | P}}
\def\A{\Delta}
\def\cano{{\bf 1}}
\def\Uv{U_{v \k \coR}}
\def\Vv{V_{v \k \coR}}
\def\x{\vec x}
\def\y{\vec y}
\def\z{\vec z}
\newcommand{\phivx}[1]{\phi_{v,\coR,\x,#1}}
\newcommand{\psivx}[1]{\psi_{v,\coR,\x,#1}}

\begin{document}

\title{Security of EPR-based Quantum Key Distribution}
\author{Hitoshi Inamori \\ Centre for Quantum Computation, Oxford University}
\date{\today}
\maketitle

\begin{abstract}
We propose a proof of the security of EPR-based quantum key distribution against enemies with unlimited computational power. The proof holds for a protocol using interactive error-reconciliation scheme. We assume in this paper that the legitimate parties receive a given number of single photon signals and that their measurement devices are perfect.
\end{abstract}


\section{Introduction}

Quantum key distribution is a cryptographic task that uses properties of quantum mechanics to allow two legitimate parties to share a secret random number. This random number can be used as a key for a symmetric classical cipher to establish a perfectly secure communication channel between the legitimate parties. The first quantum key distribution protocol, called BB84, was proposed by Bennett and Brassard~\cite{BB84}. It was followed by other protocols, such as~\cite{Ben92, Eke91} and the security of these protocols were analysed~\cite{HE94,FGGNP97,SRSF98,Lut96PRA,DEJMPS96,CG97,Lut99a,Lut00a}. The unconditional security of quantum key distribution -- i.e. security against enemies with unlimited computational power -- was obtained by Mayers~\cite{May96, May98} for the BB84 protocol and many notions and techniques introduced in the proof are used in the present paper. Other proofs of the unconditional security of BB84 followed~\cite{LC99,Lo99,BBBMR99,SP00,Ben99}. The security of EPR-based quantum key distribution protocol proposed by Ekert, E91~\cite{Eke91,BBM92}, has also been proved in~\cite{LC99,Lo99,SP00,AB00}, and the security of entanglement-based quantum key distribution using untrusted apparatus has been proved in~\cite{MY98,May99}. In this paper, we propose another proof of the security of E91. The protocol is proved secure against enemies with unlimited computational power. However, it is assumed that both legitimate parties receive an ensemble of a given number of single photons. Furthermore we assume that the efficiency of their detection unit is one, which is far from true in any practical implementation of quantum key distribution today. The results in this paper therefore do not apply to practical implementations of EPR-based quantum key distribution. Nevertheless it is hoped that techniques employed in this paper can be generalised to prove security of practical EPR-based quantum key distribution protocols.


\section{Definition of security}

We adopt the same definition of security as described in~\cite{May98,ILM00}.

The r\^ole of key distribution between two distant legitimate parties, traditionally called Alice and Bob, is to generate a shared random number, called the \emph{private key}, that is guaranteed to be known only by the legitimate parties. A non-authorised party, traditionally called Eve, should not be able to obtain any information about the private key, whichever eavesdropping strategy she might adopt. 

However, most quantum key distribution protocols do not allow Alice and Bob to share a private key in all circumstances. It is only when some conditions are satisfied that Alice and Bob can ascertain a potential eavesdropper will only have negligible information about the key. The protocol therefore provides a \emph{validation test} that tells whether a key can be generated with unconditional privacy. A key is created only if the test is passed. Otherwise the session is abandoned. Nevertheless, as in~\cite{May98,ILM00} we will adopt the convention that when the validation test is not passed, Alice chooses a random value for the private key with uniform probability distribution. As a result, the private key is defined regardless the outcome of the validation test, but, of course, when the validation test is not passed, Bob does not share the key with Alice.

Finally, we consider families of protocols for which a parameter quantifying the amount of a resource used in a protocol characterises its security. Such parameter is called \emph{security parameter}. Usually, the higher the security parameter's value is, the higher is the level of security, but also the amount of a resource required by the protocol. We now give a formal definition of security.

A random variable will always be denoted by a bold letter, and values taken by this random variable by the corresponding plain letter. Only discrete random variables will be considered in this paper. The probability distribution of a random variable $\rv{x}$ is denoted by $\pr{x}$, i.e. $\pr{x}(x)=\Pr(\rv{x}=x)$ is the probability that $\rv{x}$ takes the value $x$. The joint distribution of two random variables $\rv{x}$ and $\rv{y}$ is denoted by $\pr{ x y}$, i.e. $\pr{x y}(x,y)=\Pr(\rv{x}=x,\rv{y}=y)$. The conditional probability of $\rv{x}$ given that $\rv{y}$ takes a value $y$ is denoted by $\pc{x}{\rv{y}=y}$ whenever $\pr{y}(y)>0$, i.e. $\pc{x}{\rv{y}=y}(x) = \Pr(\rv{x}=x|\rv{y}=y)=\frac{\pr{x y}(x,y)}{\pr{y}(y)}$, whenever $\pr{y}(y)$ is positive. Let $f$ be a function defined on the image of $\rv{x}$. When no confusion is possible, the notation $\rv{f}$ will be adopted to denote the random variable $f(\rv{x})$.

We will denote by $\rv{\k}$ the random variable giving the private key generated in a key distribution session. The key is a string of $m$ bits where $m$ is a positive integer specified by the legitimate users. That is $\rv{\k}$ takes value in $\bin^m$. Given an eavesdropping strategy chosen by Eve, we denote by $\rv{v}$ the random variable giving collectively all data Eve gets during this key distribution session. Henceforth, given the eavesdropping strategy adopted by Eve, $\rv{v}$ is called the \emph{view} of Eve, and we will denote by $\V$ the set of all values $\rv{v}$ may take.

We adopt the following definition of security for quantum key distribution protocols.

\begin{definition}
Consider a quantum key distribution protocol returning a key $\rv{\k}\in\bin^m$ regardless the outcome of the validation test, where the length of the key, $m$, is fixed and chosen by the user. We say that the protocol offers \emph{perfect privacy} if and only if:
\begin{itemize}
\item the protocol is parametrised by a parameter $N$ taking value in $\Nat$ called the security parameter, and
\item there exists a function $\epsilon \,:\,\Nat \times\Nat \rightarrow \Real^+$ such that $\epsilon(N,m)$ is vanishing exponentially as $N$ grows (i.e. there exist $\alpha>0$, $\beta>0$, $N_{min}\in \Nat$ and a function $f:\,\Nat\rightarrow \Real^+$ such that $\forall N>N_{min},\, \epsilon(N,m)< e^{-\alpha N^\beta} f(m)$)  , and
\item there exists a function $N_0\,:\,\Nat\rightarrow\Nat$ such that, for any strategy adopted by Eve,
\eqa{
\lefteqn{\forall m,\,\forall N\geq N_0(m),}\nonumber\\
& &H(\rv{\k} | \rv{v}) \geq m -\epsilon(N,m)\\
}
where $\rv{v}$ is Eve's view given her strategy, and 
\eq{
H(\rv{\k} | \rv{v})\stackrel{Def}{=}-\sum_{\k,v \st \pr{\k v}(\k,v) >0}\pr{\k v}(\k,v)\log_2 \pc{\k}{\rv{v}=v}(\k)
} 
is the Shannon entropy~\cite{Sha49,Wel88,Sti95} of the key $\rv{\k}$ given Eve's view $\rv{v}$.
\end{itemize}
\end{definition}

Another important aspect of security of key distribution protocols is the \emph{integrity} or the faithfulness of the distributed key. We must require that whatever Eve does, it is very unlikely that Alice and Bob fail to share an identical private key while the validation test is passed. However, the integrity of the protocol depends mainly on the efficiency of the error detection/correction scheme that is used. This point is discussed in Appendix, but the reader is referred for instance to~\cite{BS93} for a more complete explanation.


\section{The protocol}

We describe the quantum key distribution protocol under consideration. It is a variation~\cite{BBM92} of the protocol originally proposed in~\cite{Eke91}. The protocol is designed to use classical error-reconciliation schemes like the interactive scheme proposed in~\cite{BS93}. 

\begin{description}
\item[Protocol setup]\hfill

Alice and Bob specify:
\begin{itemize}
\item $m$, the length (in bits) of the private key to be generated.
\item $\epsilon$, the maximum threshold value for the error rate during the quantum transmission ($\epsilon < 1/4$).
\item $\tau$, a security constant such that $\frac{2\epsilon}{1-\epsilon} < \frac{2 \epsilon}{1-\epsilon} +\tau < 1$.
\item the security parameter $r$. It must be large enough so that Alice and Bob can find a binary matrix $K$ of size $m\times r$ such that any linear combination of rows of $K$ that contains at least one row of $K$ has weight greater than $\dw=\left(2\frac{\epsilon}{1-\epsilon}+\tau\right) r$ (i.e. $\min_{\x\in\bin^m\setminus{\vec 0}}(\w(\x^T K))\geq \dw$ where for any vector $\y$, $\w(\y)$ is the weight of $\y$, that is, the number of non zero entries in $\y$). Alice and Bob choose one such matrix $K$. Shannon's coding theorem~\cite{Wel88} tells that for asymptotic values of $m$, such matrix can be found if $r$ obeys the inequality:
\eq{
\frac{m}{r} \leq 1 - h\left(\frac{\epsilon}{1-\epsilon}+\frac{\tau}{2}\right),
}
where $h(\epsilon)=-\epsilon \log_2 \epsilon -(1-\epsilon)\log_2 (1-\epsilon)$ is Shannon's binary entropy.
\item An error reconciliation scheme between Alice and Bob such that:
\begin{itemize}
\item it tells, with high probability of correctness, whether more than $\epsilon s$ errors are present in a string of $s$ bits, where $s=\left\lfloor \frac{r}{1-\epsilon} \right\rfloor$,
\item if there are less than $\epsilon s$ errors in the string, the scheme corrects these errors, at least with high probability of success,
\item only positions of the errors are possibly disclosed publicly. In particular the scheme should disclose no information about parities of the reconciled string.
\end{itemize}
The error reconciliation can be a probabilistic scheme for which an upper-bound on the probability of failure can be specified by Alice and Bob. One can achieve such a task by first estimating the error rate on a small randomly chosen proportion of the string and then by using for instance the interactive error-reconciliation scheme proposed in~\cite{BS93}. In these processes, the exchanged parities or bits should be encrypted with the one-time pad method~\cite{Lut96PRA,Lut99a}. A basic explanation of this scheme can be found in Appendix A, but the reader is referred to~\cite{BS93} for a complete description. The above requires that Alice and Bob share beforehand a secret private key for the one-time pad encryption. According to Shannon's coding theorem, for asymptotic values of $s$, such probabilistic error-reconciliation is possible if the entropy\footnote{That is, the length of the previously shared key if it is uniformly distributed.} $q$ (in bits) of the previously shared private key obeys the inequality:
\eq{
q \geq s h(\epsilon).
}
\item $n$, the number of pairs of photons to be sent to the legitimate parties. A good choice for $n$ is $\left\lceil\frac{r}{\frac{1-\epsilon}{2}-\tau_S}\right\rceil$ where $\tau_S$ is a small but strictly positive constant.
\end{itemize}
\item[Quantum transmission] \hfill
\begin{itemize}
\item A source sends a sequence of $n$ photons to Alice and another sequence of $n$ photons to Bob. It is assumed that ideally, for each $i\in\N$, the source emits a pair of photons in the state:
\eq{
|\phip\rangle = \frac{\ket{0}{+}\ket{0}{+}+\ket{1}{+}\ket{1}{+}}{\sqrt{2}}
}
and that Alice's $i$-th photon is the first photon of this pair, and Bob's $i$-th photon is the second photon of this pair. The kets $\ket{0}{+}$ and $\ket{1}{+}$ form an orthonormal basis $+$ of the Hilbert space describing the polarisation of one photon. The kets $\ket{0}{\times}=\frac{\ket{0}{+}+\ket{1}{+}}{\sqrt{2}}$ and $\ket{1}{\times}=\frac{\ket{0}{+}-\ket{1}{+}}{\sqrt{2}}$ form its conjugate basis $\times$.

However, the source needs not to be trusted. In particular, it can be under control of a possible eavesdropper. The only assumption is that Alice and Bob receive a sequence of $n$ single photon signals on each side.
\item We assume that the measurement devices of Alice and Bob have efficiency one. For each $i\in\N$,
\begin{enumerate}
\item Alice picks randomly a basis $a_i\in\bases$ with uniform probability distribution. Alice measures her $i$-th photon in the basis $a_i$, obtaining the outcome $\alpha_i\in\bin$, corresponding to the state $\ket{\alpha_i}{{a_i}}$.
\item Similarly, Bob picks randomly and independently of Alice a basis $b_i\in\bases$ with uniform probability distribution. Bob measures his $i$-th photon in the basis $b_i$, obtaining the outcome $\beta_i\in\bin$, corresponding to the state $\ket{\beta_i}{{b_i}}$.
\end{enumerate}
\end{itemize}
\item[Sifting] \hfill

We denote by $\a=(a_1,a_2,\ldots,a_n)$, $\b=(b_1,b_2,\ldots,b_n)$, $\aa=(\alpha_1,\alpha_2,\ldots,\alpha_n)$ and $\bb=(\beta_1,\beta_2,\ldots,\beta_n)$ the outcomes of the quantum transmission. For any vector $\vec y=(y_1,y_2,\ldots,y_n)$ in $\bin^n$ or $\bases^n$, and for any subset $X$ of $\N$, we denote by $y_X$ the restriction of $\vec y$ on $X$.

Alice and Bob compare publicly their bases $\a$ and $\b$. We denote by $\d$ the vector in $\bin^n$ such that for any $i\in\N$, $d_i=1$ if and only if $a_i=b_i$. If the number of indexes $i\in\N$ such that $a_i=b_i$ is greater than or equal to $s$ (i.e. $\w(\d)\geq s$) then the \emph{sifted set} $S$ is defined as the set of the first $s$ such indexes. Otherwise the validation test is failed. The bit strings $\aaS$ and $\bbS$ are usually referred to as the \emph{sifted keys}.

\item[Error correction]\hfill

Alice and Bob perform the error correction on their sifted keys $\aaS$ and $\bbS$ as specified in the protocol setup.  We define the \emph{error set} $E$ as the set of indexes in $S$ in which an error is found, that is, $\alpha_i\neq\beta_i$. Likewise, we define the \emph{error vector} $\e$ as the vector in $\bin^s$ giving the positions of the errors ($\forall i\in\{1,\ldots,s\}$, $e_i=1$ if and only if $\alpha_i \neq \beta_i$). We denote by $e$ the size of the set $E$, i.e. $e=|E|=\w(\e)$. The validation test is passed if $e < \epsilon s$, otherwise it is failed. If the validation test is passed, then Alice and Bob define the \emph{reconciled set} $R$ as the set of the first $r$ indexes $i\in S\setminus E$ \footnote{ Note that $|S \setminus E|\geq r$ if the validation test is passed.}.  Therefore $|R|=r$ and $\forall i\in R$, $a_i=b_i$ and $\alpha_i=\beta_i$. Alice and Bob obtain an identical string of bits $\aaR\in\bin^r$, called the \emph{reconciled key}.
\item[Privacy amplification]\hfill

The private key is defined as:
\begin{enumerate}
\item $\k = K \aaR \pmod{2}$ if the validation test is passed.
\item an $m$-bit string $\k$ picked randomly by Alice with uniform probability distribution each time the validation test is failed.
\end{enumerate}
\end{description}


\section{Privacy of the protocol}
The main result of this paper is stated.
\begin{property}
The protocol described above offers perfect privacy: for any eavesdropping strategy chosen by a possible eavesdropper, the conditional entropy of the private key $\rv{\k}$ given the eavesdropper's view $\rv{v}$ is bounded from below by:
\eq{
H(\rv{\k} | \rv{v}) \geq m - 2\left(m+\frac{1}{\ln 2}\right)\left(\theta(r)+2\sqrt{\theta(r)}\right),
}
where
\eq{
\theta(r) = 2^{-\left(1-h\left(\frac{1}{2}-\frac{3}{16}\tau\right)\right)\frac{\tau}{2} r}.
}

The above bound applies for any value of the security parameter $r$ such that the matrix $K$ specified in the protocol exists.
\end{property}

The protocol uses previously shared private key for the error reconciliation. 
A net gain in shared private bits is achieved if $m$ is greater than the number of the secret bits used during error reconciliation. For asymptotic values of $m$, we can take arbitrarily small values for the security parameter and a net gain in private bits is obtained if $m > s h(\epsilon) \simeq \frac{r}{1-\epsilon} h(\epsilon)$. We have seen that privacy amplification is possible if $\frac{m}{r} \leq 1-h\left(\frac{\epsilon}{1-\epsilon}\right)$. Therefore, a net gain in shared private bits can be obtained for asymptotic values of $m$ if:
\eq{
1-h\left(\frac{\epsilon}{1-\epsilon}\right)-\frac{1}{1-\epsilon}h(\epsilon) > 0.
}


\section{Proof of the privacy}


\subsection{Notations}

We define the notations used throughout the proof.

\begin{description}

\item[Classical data] \hfill

We denote collectively by $C=(\a,\b,\aa,\bb)$ the classical data Alice and Bob generate during the protocol (after the setup). Note that other variables Alice and Bob generate during the protocol can be deterministically derived from $C$. We denote by $P=(\a,\d,\e)$ the data that are publicly announced by Alice and Bob during the protocol. Recall that specifying $\a$ and $\d$ is equivalent to specifying $\a$ and $\b$. For any possible $P$, we denote by $\C_P$ the set of values for the classical data that are compatible with the public announcement of $P$. That is, for a given $P=(\a,\d,\e)$, 
\eqa{
\C_P &=& \{ C'=(\a',\b',\aa',\bb') \st \a'=\a,\nonumber\\
& &\forall i,\, b'_i=a_i \mbox{ if } d_i=1,\, b'_i \neq a_i \mbox{ if } d_i=0\nonumber\\
& &\forall i\in E,\,\alpha'_i \neq \beta'_i \mbox{ and } \forall i\in S\setminus E,\,\alpha'_i = \beta'_i \\
& &\mbox{ where $S$ and $E$ are given by $\d$ and $\e$.} \}.
} 

Given a possible $P$ and a value for the private key $\k$, we define $\C_{P,\k}$ as the set of values for the classical data that are compatible with the public announcement of $P$ and generation of $\k$ for the private key. That is, for a given $P=(\a,\d,\e)$, 
\eqa{
\C_{P,\k} &=& \{ C'=(\a',\b',\aa',\bb') \st \a'=\a,\\
& &\forall i,\, b'_i=a_i \mbox{ if } d_i=1,\, b'_i \neq a_i \mbox{ if } d_i=0\nonumber\\
& &\forall i\in E,\,\alpha'_i \neq \beta'_i \mbox{ and } \forall i\in S\setminus E,\,\alpha'_i = \beta'_i \\
& &K \aaR' = \k\pmod{2}, \\
& &\mbox{ where $S$, $E$ and $R$ are given by $\d$ and $\e$.} \}.
} 

Finally, we denote by $\P$ the set of all possible public announcements for which the validation test is passed. That is, 
\eq{
\P = \{ P=(\a,\d,\e) \st \w(\d)\geq s \mbox{ and } e < \epsilon s\}.
}

For any vector $\vec x$ and any symbol $A$, $\w(\vec x)$ is the number of non-zero entries, and $\w_A(\vec x)$ is the number of entries with symbol $A$. For any vector $\x\in\bin^n$, we denote by $\neg\x$ the vector whose $i$-th entry is $1+x_i\pmod{2}$ for all $i\in\N$. 
Finally, we denote by $T$ the subset $S\setminus(E\cup R)$, and by $t$ the size of $T$.

\item[Bell states] \hfill

For each $i\in\N$, we define the Bell basis $\{\ket{0}{i},\,\ket{1}{i},\,\ket{2}{i},\,\ket{3}{i}\}$ of the $i$-th pair of photons as:
\eqa{
\ket{0}{i} &=& \frac{\ket{0}{+,i}\ket{0}{+,i}+\ket{1}{+,i}\ket{1}{+,i}}{\sqrt{2}},\\
\ket{1}{i} &=& \frac{\ket{0}{+,i}\ket{0}{+,i}-\ket{1}{+,i}\ket{1}{+,i}}{\sqrt{2}},\\
\ket{2}{i} &=& \frac{\ket{0}{+,i}\ket{1}{+,i}+\ket{1}{+,i}\ket{0}{+,i}}{\sqrt{2}},\\
\ket{3}{i} &=& \frac{\ket{0}{+,i}\ket{1}{+,i}-\ket{1}{+,i}\ket{0}{+,i}}{\sqrt{2}},
}
where the first and the second state in the product states in the rhs.~correspond to Alice's and Bob's $i$-th photon's polarisation state, respectively. Tensor products are implied when we consider state of several photons, that is, $\ket{\aa,\bb}{\a,\b}=\otimes_{i=1}^n \ket{\alpha_i}{a_i,i}\ket{\beta_i}{b_i,i}$ and $\ketb{\c}=\otimes_{i=1}^n \ket{c_i}{i}$. For any subset $X$ of $\N$, $\ket{\alpha_{\ens X},\beta_{\ens X}}{a_{\ens X},b_{\ens X}}=\otimes_{i\in X} \ket{\alpha_i}{a_i,i}\ket{\beta_i}{b_i,i}$.

Given a basis $a\in\bases$, we define $X_a$ as the set of indexes of Bell states that are compatible with Alice and Bob measuring in the same basis $a$ and sharing the same bit value. Likewise, we define $Y_a$ as the set of indexes of Bell states that are compatible with Alice and Bob measuring in basis $a$ and not sharing the same bit value. That is, $X_+ = \{0,1\}$, $X_{\times} = \{0,2\}$, $Y_+ = \{2,3\}$ and $Y_{\times} = \{1,3\}$. Given the choice of bases $\a$ and a set $A\subset\N$, we define $X_{a_{\ens A}}$ as $\{ c_{\ens A}\in\bell^A \st \forall i\in A, c_i\in X_{a_i}\}$ and $Y_{a_{\ens A}}$ as $\{ c_{\ens A}\in\bell^A \st \forall i\in A, c_i\in Y_{a_i}\}$. Given a reconciled set $R$ and the choice of bases $\aR$ on $R$, for any $\cR\in X_{\aR}$, we will denote by $\cc$ the unique $\cc\in\bin^r$ such that for each $i\in\{1,\ldots,r\}$, $c_i=(1+\w_\times a_i)\gamma_i$, i.e. $c_i=\gamma_i$ if $a_i=+$, and $c_i=2\gamma_i$ if $a_i=\times$. For any vectors $\vec x$, $\vec y$ $\in\bin^r$, we define $\x \scal\y$ as $\vec x \scal \vec y \stackrel{Def}{=} \sum_{i=1}^r x_i y_i$. Given $R$ and $\aR$,  for any $\cR\in X_{\aR}$, we have the identity $\bra{\aaR,\aaR}{\aR} \cR \rangle = \frac{(-1)^{\aaR\scal\cc}}{\sqrt{2}^r}$.

\end{description}


\subsection{Model of measurements}

A mathematical model of measurements on the quantum state generated by the source is given. 
The source can be under complete control of Eve, as long as it sends $n$ single photons to both Alice and Bob. In such a scenario, Eve may entangle a probe of any dimension to the photons she sends to Alice and Bob which are in any state Eve wants. We write the state of these $n$ couples of photons and the probe in the Bell basis as follows:
\eq{
\rho = \sum_{\c,\c'}\ketb{E_{\c}}\brab{E_{\c'}}\otimes\ketb{\c}\brab{\c'},
}
where the states $\ketb{E_{\c}}$ are states of Eve's probe that are possibly nor orthogonal nor normalised.
The positive operator giving the probability that Alice and Bob get $C=(\a,\b,\aa,\bb)$ as their classical data is simply:
\eq{
F_C=\pr{\a}(\a)\pr{\b}(\b)\proj{\aa,\bb}{\a,\b},
}
where $\pr{\a}(\a)=1/2^n$ and $\pr{\b}(\b)=1/2^n$ for any choice of $\a$ and $\b$. Note that $\pr{\a}(\a)\pr{\b}(\b)=\pr{\a}(\a)\pr{\d}(\d)$ where $\pr{\d}=1/2^n$.

The positive operator giving the probability that Alice and Bob publicly announce $P=(\a,\d,\e)$ while they get the private key $\k$ is the sum of the operators $F_{C'}$ for $C'$ running over $\C_{P,\k}$:
\eqa{
F_{P,\k}&=&\sum_{C'\in\C_{P,\k}}\pr{\a}(\a')\pr{\d}(\d')\proj{\aa',\bb'}{\a',\b'}\\
&=&\pr{\a}(\a)\pr{\d}(\d)\Id_{\overline{S}}\otimes\sum_{\aaE\in\bin^e}\proj{\aaE,\neg\aaE}{\aE,\aE}\nonumber\\
& &\otimes \sum_{\aaT\in\bin^t}\proj{\aaT,\aaT}{\aT,\aT}\nonumber\\
& &\otimes\sum_{\begin{array}{l}\scriptstyle \aaR\in\bin^r\st\\ \scriptstyle K\aaR = \k\end{array}}\proj{\aaR,\aaR}{\aR,\aR}
}
where $\Id_{\overline{S}}$ is the identity operator acting on the Hilbert space describing photons not in $S$. Note that $\b(S)=\a(S)$.

Similarly, the positive operator giving the marginal probability that Alice and Bob publicly announce $P=(\a,\d,\e)$ is the sum of the operators $F_{C'}$ for $C'$ running over $\C_P$:
\eqa{
F_{P}&=&\sum_{C'\in\C_P}\pr{\a}(\a')\pr{\d}(\d')\proj{\aa',\bb'}{\a',\b'}\\
&=&\pr{\a}(\a)\pr{\d}(\d)\Id_{\overline{S}}\otimes\sum_{\aaE\in\bin^e}\proj{\aaE,\neg\aaE}{\aE,\aE}\nonumber\\
& &\otimes \sum_{\aaT\in\bin^t}\proj{\aaT,\aaT}{\aT,\aT}\nonumber\\
& &\otimes\sum_{\aaR\in\bin^r}\proj{\aaR,\aaR}{\aR,\aR}.
}

Eve may perform a general measurement on her probe. This general measurement can take place after Alice and Bob's public announcements and therefore can be conditioned on $\rv{P}$. We will denote by $\V_{P}$ the set of views $v$ that are compatible with the public announcement $P$. The positive operator giving the probability that Eve gets the view $v$ given that Alice and Bob announced $P$ will be denoted by $G_{v | P}$. We will assume without loss of generality that the operators $G_{v | P}$ are of rank one, i.e. $G_{v|P} = \projb{\phiv}$ where the vectors $\ketb{\phiv}$ are possibly not orthogonal nor normalised, but obey the relation $\sum_{v\in\V_{P}} G_{v|P} = \Id$.


\subsection{The r\^ole of the validation test}

Here we show that it is very unlikely that the validation test is passed when the state of the photons emitted by the source is very different from the ideal state specified by the protocol. The underlying principle has been advanced in~\cite{LC99,SP00,Ben99}. More precisely, given a possible reconciled set $R$, let $\Pi_R$ be the orthogonal projection operator defined as:
\eqa{
\Pi_R &=& \sum_{\begin{array}{l} \scriptstyle\c\in\bell^n\st \\ \scriptstyle\w(\cR) \geq \dw/2\end{array}} \projb{\c}\\
&=& \Id_{\overline{R}} \otimes \sum_{\begin{array}{l}\scriptstyle\cR\in\bell^r\st\\ \scriptstyle\w(\cR)\geq \dw/2 \end{array}}\projb{\cR}.
} 

The operator $\Pi_R$ projects onto Bell states for pairs of photons in $R$ with weight greater than $\dw/2=\left(\frac{\epsilon}{1-\epsilon}+\frac{\tau}{2}\right)r$. The following property is then proved.

\begin{property}
The eigenvalues of the semi-definite positive Hermitian operator
\eq{
\sum_{P\in \P} \Pi_R F_P \Pi_R,
}
where $R$ is specified by $P$ in the sum, are bounded from above by
\eq{
\theta(r) = 2^{-\left(1-h\left(\frac{1}{2}-\frac{3}{16}\tau\right)\right)\frac{\tau}{2} r}.
}
\end{property}

\proof 
The above operator can be written as:
\eq{
\sum_{P\in \P} \Pi_R F_P \Pi_R = \sum_{\begin{array}{l}\scriptstyle\d\in\bin^n\st\\ \scriptstyle \w(\d) \geq s\end{array}}\sum_{\begin{array}{l}\scriptstyle\e\in\bin^s\st\\ \scriptstyle \w(\e) <\epsilon s\end{array}} \Pi_R \left(\sum_{\a} F_P\right)\Pi_R.
}
Now for given $\d$ and $\e$,
\eq{
\sum_{\a} F_P = \pr{\d}(\d) \Id_{\overline{S}} \otimes_{i\in E} Y_i \otimes_{j\in T} X_j \otimes_{k\in R} X_k,
}
where
\eqa{ 
X_i &=&  \projb{0}+\frac{1}{2} \projb{1}+\frac{1}{2}\projb{2}, \\
Y_i &=&  \projb{3}+\frac{1}{2}\projb{1}+\frac{1}{2}\projb{2}
}
are operators acting on $i$-th photon pair's Hilbert space. The last equalities are derived directly from the definition of the Bell states. As a consequence, we have,
\eq{
\Pi_R \left(\sum_{\a} F_P\right) \Pi_R = \pr{\d}(\d) \Id_{\overline{S}}\otimes_{i\in E} Y_i \otimes_{j\in T}X_j\otimes\Big(\sum_{\begin{array}{l} \scriptstyle \cR \in\{0,1,2\} \st\\ \scriptstyle \w(\cR)\geq \dw/2\end{array}} \frac{\projb{\cR}}{2^{\w_1(\cR)+\w_2(\cR)}}\Big). 
}

Now, given $\d\in\bin^n$, the operator:
\eq{
\sum_{\e\st \w(\e)<\epsilon s} \Pi_R \left(\sum_{\a} F_P\right) \Pi_R
}
is diagonal in the Bell basis $\ketb{c}$. Given a vector $\c\in\bell^n$ and an error vector $\e\in\bin^s$, a necessary condition for the scalar:
\eq{
\brab{c}\Pi_R \left(\sum_{\a} F_P\right) \Pi_R\ketb{c}
}
to be non zero is that, for all $i\in S$,
\begin{itemize}
\item $e_i=0$ if $c_i=0$,
\item $e_i=1$ if $c_i=3$, and
\item $\w_1(\cS)+\w_2(\cS) \geq e-\w_3(\cS)+\dw/2$ (otherwise $\w(\cR)$ is smaller than $\dw/2$).
\end{itemize}
Let $k=e-\w_3(\cS)$. Then there are $\binom{\w_1(\cS)+\w_2(\cS)}{k}$ such vectors $\e$ of weight $e$, if $0\leq k <\epsilon s-\w_3(\cS)$ and $k\leq \w_1(\cS)+\w_2(\cS)-\dw/2$. Therefore,
\eqa{
\lefteqn{\brab{c}\sum_{\e\st \w(\e)<\epsilon s}\Pi_R \left(\sum_{\a} F_P\right) \Pi_R\ketb{c}}\nonumber\\
&\leq& \frac{\pr{\d}(\d)}{2^{\w_1(\cS)+\w_2(\cS)}}\sum_{\begin{array}{l}\scriptstyle 0\leq k < \epsilon s-\w_3(\cS),\,\mbox{\scriptsize and}\\\scriptstyle k\leq \w_1(\cS)+\w_2(\cS)-\frac{\dw}{2}\end{array}}\binom{\w_1(\cS)+\w_2(\cS)}{k}.
}

Now, $\dw$ is either greater or smaller than $(\w_1(\cS)+\w_2(\cS))\left(1+\frac{\tau}{2}(1-\epsilon)\right)$.
\begin{itemize}
\item If $\dw > (\w_1(\cS)+\w_2(\cS))\left(1+\frac{\tau}{2}(1-\epsilon)\right)$, then
\eq{
\w_1(\cS)+\w_2(\cS)-\frac{\dw}{2} < \frac{1}{2}\left(1-\frac{\tau}{2}(1-\epsilon)\right)(\w_1(\cS)+\w_2(\cS)) \,\mbox{ and,}
}
\item if $\dw \leq (\w_1(\cS)+\w_2(\cS))\left(1+\frac{\tau}{2}(1-\epsilon)\right)$, then
\eqa{
\epsilon s-\w_3(\cS) &\leq& \frac{\epsilon r}{1-\epsilon}\\
&=& \frac{\dw}{2}-\frac{\tau}{2} r\\
&\leq& \frac{1}{2}\left(1-\frac{\tau}{2}(1-\epsilon)\right)(\w_1(\cS)+\w_2(\cS)),
}
where we have used $r\geq s(1-\epsilon)$ and $s\geq \w_1(\cS)+\w_2(\cS)$.
\end{itemize}
We thus derived that:
\eqa{
\lefteqn{\brab{c}\sum_{\e\st \w(\e)<\epsilon s}\Pi_R \left(\sum_{\a} F_P\right) \Pi_R\ketb{c}}\nonumber\\
&\leq& \frac{\pr{\d}(\d)}{2^{\w_1(\cS)+\w_2(\cS)}}\sum_{0\leq k <\frac{1}{2}\left(1-\frac{\tau}{2}(1-\epsilon)\right)(\w_1(\cS)+\w_2(\cS)) }\binom{\w_1(\cS)+\w_2(\cS)}{k}\\
&\leq& \pr{\d}(\d) 2^{-\left(1-h\left(\frac{1}{2}\left(1-\frac{\tau}{2}(1-\epsilon)\right)\right)\right)(\w_1(\cS)+\w_2(\cS))}\\
&\leq& \pr{\d}(\d) 2^{-\left(1-h\left(\frac{1}{2}-\frac{3}{16}\tau\right)\right)\frac{\tau}{2}r}\\
&=& \pr{\d}(\d) \theta(r)
}
where we have used the binomial inequality stating that $\sum_{0\leq k < p n} \binom{n}{k} \leq 2^{n h(p)}$ for any positive integer $n$ and $0\leq p < 1/2$. In the last inequality we have used the inequalities $\epsilon <1/4$ and $\w_1(\cS)+\w_2(\cS)\geq\dw/2$ when the above scalar is non zero.

Remarking that the operator $\sum_{\a,\e\st\ e<\epsilon s}\Pi_R F_P\Pi_R$ is diagonal in the Bell basis for all $\d$ and $\sum_{\d \st \w(\d) \geq s} \pr{\d}(\d) \leq 1$, this concludes the proof.\fin

We recall that $\rho = \sum_{\c,\c'}\ketb{E_{\c}}\brab{E_{\c'}}\otimes\ketb{\c}\brab{\c'}$ is the density operator describing Alice and Bob's photons and Eve's probe. The above property implies that:
\eq{
\Tr\Big(\Id_{\mbox{\scriptsize Eve}}\otimes\sum_{P\in\P} \Pi_R F_P \Pi_R \rho\Big) \leq \theta(r)
}
where $\Id_{\mbox{\scriptsize Eve}}$ is the identity operator acting on the Hilbert space of the probe. That is,
\eq{
\sum_{P\in\P}\pr{\a}(\a)\pr{\d}(\d) \sum_{\begin{array}{l}\scriptstyle \coR \st \\\scriptstyle \cE\in Y_{\aE},\\\scriptstyle\cT\in X_{\aT}\end{array}}\sum_{\begin{array}{l}\scriptstyle\cR\in X_{\aR}\st\\\scriptstyle\w(\cR) \geq \dw/2\end{array}} \scalarb{E_{\c}}{E_{\c}} \leq \theta(r).
}


\subsection{Quasi-independence of the key and the view}

In this section we compute the joint probability distribution of the key and the view. We prove that this distribution is very close to a product of an uniform distribution for the key and the marginal probability distribution of the view.

\begin{property}
For any given eavesdropping strategy chosen by Eve and returning a view $\rv{v}$, the probability distribution of the key $\rv{\k}$ and the view $\rv{v}$ obeys the following inequality:
\eq{
\sum_{P\in\P}\sum_{v\in\V_P}\sum_{\k\in\bin^m} \left|\pr{\k v}(\k, v)-\frac{1}{2^m}\pr{v}(v)\right | \leq 2\left(\theta(r)+2\sqrt{\theta(r)}\right)
}
where $m$ is the length of the private key and $r$ is the size of the reconciled set.
\end{property}

\proof For any $\k\in\bin^m$, $P$ and $v\in\V_P$, we have:
\eqa{
\lefteqn{\pr{\k v}(\k,v) - \frac{1}{2^m}\pr{v}(v)}\nonumber\\
&=& \Tr(G_{v | P}\otimes F_{P,\k}\, \rho) - \frac{1}{2^m} \Tr(G_{v|P}\otimes F_P\,\rho) \\
&=& \pr{\a}(\a)\pr{\d}(\d)\sum_{\begin{array}{l}\scriptstyle \c,\c'\st\\ \scriptstyle \cE,\cE'\in Y_{\aE},\\ \scriptstyle \cT,\cT'\in X_{\aT},\\ \scriptstyle \cR,\cR'\in X_{\aR}\end{array}} \brab{E_{\c'}} G_{v|P} \ketb{E_{\c}}\delta_{\coR,\coR'} d_{\k}(\cc,\cc'),
}
where
\eq{
d_{\k}(\cc,\cc') =  \sum_{\begin{array}{l}\scriptstyle \aaR\in\bin^r\st\\ \scriptstyle K\aaR = \k\pmod{2}\end{array}} \frac{(-1)^{\aaR \scal (\cc+\cc')}}{2^r} - \frac{1}{2^m}\delta_{\cc,\cc'}.
}
where we have used the identity $\bra{\aaR,\aaR}{\aR,\aR} \cR\rangle = \frac{(-1)^{\aaR\scal\cc}}{\sqrt{2}^r}$ for any $\cR\in X_{\aR}$. We denote by:
\begin{itemize}
\item $\genK$ the set of all linear combinations over $\bin$ of rows of $K$. It is a vector space of dimension $m$.
\item $\suppl$ a subspace of $\bin^r$ that is supplement to the subspace $\genK$, that is $\genK\oplus\suppl=\bin^r$. The dimension of $\suppl$ is $r-m$.
\item $\kerK$ the set of all vectors $\vec x \in\bin^r$ such that $K \vec x = \vec 0 \pmod{2}$. The set $\kerK$ is a vector space of dimension $r-m$, since the rows of $K$ are linearly independent. We will denote by $\{\vec u_1,\ldots,\vec u_{r-m}\}$ a basis of $\kerK$. 
\end{itemize}

Given a subspace $F$ of $\bin^r$, we denote by $F^\ortho$ the set of all vectors $\vec x \in\bin^r$ such that for all $\vec y \in F$, $\vec x \scal \vec y=0 \pmod{2}$. Remark that $\kerK^\ortho=\genK$. Since the rows of $K$ are linearly independent, for any $\k\in\bin^m$, there exists a vector $\ttheta \in \bin^r$ such that $K\ttheta=\k\pmod{2}$. It follows that $K\aaR=\k\pmod{2}$ if and only if $\aaR \in \ttheta + \kerK$. Thus following the techniques used in~\cite{May98},
\eqa{
\sum_{\begin{array}{l}\scriptstyle \aaR\in\bin^r\st\\ \scriptstyle K\aaR = \k\pmod{2}\end{array}} (-1)^{\aaR \scal (\cc+\cc')} 
&=& \sum_{\aaR\in\ttheta+\kerK} (-1)^{\aaR\scal(\cc+\cc')} \\
&=& (-1)^{\ttheta\scal(\cc+\cc')}\prod_{i=1}^{r-m} \left[1+(-1)^{\vec u_i \scal (\cc+\cc')}\right] \\
&=& \left\{ \begin{array}{ll} (-1)^{\ttheta\scal(\cc+\cc')}2^{r-m} & \,\mbox{ if }\, \cc+\cc' \in \kerK^\ortho=\genK, \\ 0 & \,\mbox{ if }\, \cc+\cc'\notin\genK. \end{array}\right. 
}

One obtains therefore that:
\eq{
\pr{\k v}(\k,v) - \frac{1}{2^m}\pr{v}(v) = \frac{1}{2^m} \pr{\a}(\a)\pr{\d}(\d) \sum_{\begin{array}{l}\scriptstyle \coR\st\\ \scriptstyle \cE\in Y_{\aE},\\\scriptstyle\cT\in X_{\aT}\end{array}} (\Uv+\Vv)^\dag \A (\Uv+\Vv),
}
where $\Uv$ and $\Vv$ are complex vectors of dimension $2^r$ and $\A$ is a $2^r\times 2^r$ complex matrix, whose entries are indexed by $\cc\in\bin^r$. The $\cc$-th entry of $\Uv$ and $\Vv$ are:
\eqa{
\Big( \Uv \Big)_{\cc} &=& \left\{\begin{array}{cl}(-1)^{\ttheta\scal\cc}\brab{\phiv} E_{\c} \rangle & \,\mbox{ if }\, \w(\cc) < \dw / 2, \\ 0 &\,\mbox{ if }\, \w(\cc) \geq \dw/2. \end{array}\right.\\
\Big( \Vv \Big)_{\cc} &=& \left\{\begin{array}{cl}0 &\,\mbox{ if }\, \w(\cc) < \dw/2,\\(-1)^{\ttheta\scal\cc}\brab{\phiv} E_{\c} \rangle & \,\mbox{ if }\, \w(\cc) \geq \dw / 2, \\ \end{array}\right.
}
where $\c$ is given by $\coR$, $\aR$ and $\cc$. The $(\cc,\cc')$-th entry of $\A$ is:
\eq{
\Big( \A \Big)_{\cc, \cc'} = \left\{ \begin{array}{ll} 1 &\,\mbox{ if }\, \cc+\cc'\in\genK\setminus \{0\},\\ 0 &\,\mbox{ if }\, \cc+\cc'\notin\genK\setminus \{0\}.\end{array}\right.
}

This implies $\Uv^\dag\A\Uv=0$, since $\w(\cc) <\dw/2$ and $\w(\cc') < \dw/2$ imply that $\w(\cc+\cc') < \dw$, that is, $\cc+\cc'\notin\genK\setminus\{\vec 0\}$.

The matrix $\A$ is Hermitian, of eigenvalues $2^m-1$ and $-1$. There are $2^{r-m}$ eigenvectors $\vec v_{\vec x}$ ($\vec x\in\suppl$) associated with the eigenvalue $2^m-1$. The $\cc$-th entry of $\vec v_{\vec x}$ is:
\eq{
\Big(\vec v_{\vec x}\Big)_{\cc} = \left\{\begin{array}{cl} 1 &\,\mbox{ if }\, \cc+\vec x\in\genK \\ 0 &\,\mbox{ if }\, \cc+\vec x\notin\genK.\end{array}\right.
}

There are $2^{r-m}\left(2^m-1\right)$ eigenvectors $\vec w_{\vec x,\vec \sigma}$ ($\vec x\in\suppl$, $\vec \sigma\in\bin^m\setminus\{\vec 0\}$) associated with the eigenvalue $-1$. The $\cc$-th entry of $\vec w_{\vec x,\vec \sigma}$ is:
\eq{
\Big(\vec w_{\vec x,\vec \sigma}\Big)_{\cc} = \left\{ \begin{array}{cl} (-1)^{\vec\omega_{\cc+\vec x}\scal \vec \sigma} & \,\mbox{ if }\, \cc+\vec x\in\genK \\ 0 & \,\mbox{ if }\, \cc+\vec x\notin\genK.\end{array}\right.
}
where for any $\vec y \in \genK$, $\vec\omega_{\vec y}$ is the unique vector in $\bin^m$ such that $K^T\vec\omega_{\vec y}=\vec y\pmod{2}$. Note that for any $\cc\in\bin^r$, there is an unique $(\vec x,\vec y)\in\suppl\times\genK$ such that $\cc=\vec x+\vec y$, and that:
\eq{
\cano_{\cc} = \frac{1}{2^m}\left(\vec v_x + \sum_{\vec \sigma \in\bin^m\setminus\{\vec 0\}} (-1)^{\vec\omega_{\vec y}\scal\vec \sigma} \vec w_{\vec x, \vec \sigma}\right),
}
where $\cano_{\cc}$ is the canonical vector with entry $1$ at position $\cc$ and $0$ everywhere else. We can express the vectors $\Uv$ and $\Vv$ as linear combinations of these eigenvectors:
\eqa{
\Uv &=& \sum_{\x\in\suppl} (-1)^{\ttheta\scal\x} \phivx{\k} \vec v_{\x} + \sum_{\x\in\suppl}\sum_{\vec \sigma \neq \vec 0} (-1)^{\ttheta\scal\x} \phivx{\k+\vec \sigma} \vec w_{\x, \vec \sigma}, \\
\Vv &=& \sum_{\x\in\suppl} (-1)^{\ttheta\scal\x} \psivx{\k} \vec v_{\x} + \sum_{\x\in\suppl}\sum_{\vec \sigma \neq \vec 0} (-1)^{\ttheta\scal\x} \psivx{\k+\vec \sigma} \vec w_{\x, \vec \sigma},
}
where for any $\z\in\bin^m$,
\eqa{
\phivx{\z} &=& \sum_{\begin{array}{l}\scriptstyle\y\in\genK\st\\\scriptstyle \w(\x+\y) < \dw/2\end{array}}   \frac{(-1)^{\vec\omega_{\y}\scal\z}}{2^m}\scalarb{\phiv}{E_{\c}},\\
\psivx{\z} &=& \sum_{\begin{array}{l}\scriptstyle\y\in\genK\st\\\scriptstyle \w(\x+\y) \geq \dw/2\end{array}}\frac{(-1)^{\vec\omega_{\y}\scal\z}}{2^m}\scalarb{\phiv}{E_{\c}}.
}

In deriving the above formulae, we used the identity $\ttheta\scal\y=\vec\omega_{\y}\scal\k\pmod{2}$ for any $\y\in\genK$ and $\k\in\bin^m$. It follows that:
\eqa{
\Vv^\dag \A\Vv &=& \sum_{\x\in\suppl} |\psivx{\k}|^2 (2^m-1) \| \vec v_{\x} \|^2 - \sum_{\begin{array}{l}\scriptstyle \x\in\suppl \\ \scriptstyle \vec \sigma\neq \vec 0 \end{array}} |\psivx{\k+\vec\sigma}|^2 \|\vec w_{\x,\vec\sigma} \|^2 \\
&=& 2^m \Big[ (2^m-1)  \sum_{\x\in\suppl} |\psivx{\k}|^2-\sum_{\begin{array}{l}\scriptstyle \x\in\suppl \\ \scriptstyle \vec \sigma\neq \vec 0 \end{array}} |\psivx{\k+\vec\sigma}|^2 \Big],
}
thus,
\eqa{
\lefteqn{\sum_{\k\in\bin^m} \Big| \Vv^\dag \A \Vv \Big|}\nonumber\\ 
&\leq& 2^m \sum_{\x\in\suppl} \Big[ (2^m-1)\sum_{\k} |\psivx{\k}|^2+\sum_{\vec\sigma\neq\vec 0,\k}|\psivx{\k+\vec\sigma}|^2 \Big]\\
&=& 2^{m+1}(2^m-1)\sum_{\x\in\suppl}\sum_{\k} |\psivx{\k}|^2.
}

Similarly, we have,
\eq{
\sum_{\k\in\bin^m} \Big| \Uv^\dag \A \Vv \Big|
\leq 2^{m+1}(2^m-1) \sum_{\x\in\suppl}\sum_{\k} |\phivx{\k}^*\psivx{\k}|.
}

Now,
\eqa{
\lefteqn{\sum_{P\in\P}\sum_{v\in\V_P}\sum_{\k\in\bin^m}\left|\pr{\k v}(\k,v) - \frac{1}{2^m}\pr{v}(v)\right|}\nonumber\\
&\leq& \sum_{P\in\P}\frac{1}{2^m} \pr{\a}(\a)\pr{\d}(\d) \sum_{\ScoR}\sum_{v\in\V_P}\sum_{\k\in\bin^m}\Big[ |\Vv^\dag \A \Vv| + 2 |\Uv^\dag\A\Vv| \Big]\\
&\leq& 2(2^m-1)\sum_{P\in\P} \pr{\a}(\a)\pr{\d}(\d) \sum_{\ScoR}\sum_{v\in\V_P} \sum_{\x\in\suppl}\sum_{\k}\Big[ |\psivx{\k}|^2+2|\phivx{\k}^*\psivx{\k}|\Big]\\
&\leq& 2(2^m-1)(\eta + 2\sqrt{\eta}\sqrt{\xi}),
}
where we used the Schwartz inequality, and where
\eqa{
\eta &=& \sum_{P\in\P}\sum_{\ScoR}\sum_{v\in\V_P}\sum_{\x\in\suppl}\sum_{\k}  \pr{\a}(\a)\pr{\d}(\d) |\psivx{\k}|^2,\\
\xi  &=& \sum_{P\in\P}\sum_{\ScoR}\sum_{v\in\V_P}\sum_{\x\in\suppl}\sum_{\k}  \pr{\a}(\a)\pr{\d}(\d) |\phivx{\k}|^2.
}

We derive an upper-bound on $\eta$ and $\xi$. We have:
\eqa{
\eta &=&  \sum_{P\in\P}\pr{\a}(\a)\pr{\d}(\d)\sum_{\ScoR}\sum_{v\in\V_P}\sum_{\x\in\suppl}\sum_{\begin{array}{l}\scriptstyle \y, \y'\in\genK \\ \scriptstyle \w(\x+\y) \geq \dw/2 \\ \scriptstyle \w(\x+\y') \geq \dw/2\end{array}}\sum_{\k} \frac{(-1)^{\vec\omega_{\y+\y'}\scal\k}}{2^{2m}}\scalarb{E_{\c'}}{\phiv}\scalarb{\phiv}{E_{\c}}\\
&=& \frac{1}{2^{m}}\sum_{P\in\P}\pr{\a}(\a)\pr{\d}(\d)\sum_{\ScoR}\sum_{\x\in\suppl}\sum_{\begin{array}{l}\scriptstyle\y\in\genK \\\scriptstyle \w(\x+\y)\geq \dw/2\end{array}}\sum_{v\in\V_P} \scalarb{E_{\c}}{\phiv}\scalarb{\phiv}{E_{\c}}\\
&=& \frac{1}{2^{m}}\sum_{P\in\P}\pr{\a}(\a)\pr{\d}(\d)\sum_{\ScoR}\sum_{\begin{array}{l}\scriptstyle\x\in\suppl,\,\scriptstyle\y\in\genK \\\scriptstyle \w(\x+\y)\geq \dw/2\end{array}}\scalarb{E_{\c}}{E_{\c}}\\
&=& \frac{1}{2^{m}}\sum_{P\in\P}\pr{\a}(\a)\pr{\d}(\d)\sum_{\ScoR}\sum_{\begin{array}{l}\scriptstyle\cR\in X_{\aR}\st\\\scriptstyle \w(\cR) \geq \dw/2 \end{array}} \scalarb{E_{\c}}{E_{\c}}\\
&\leq& \frac{1}{2^{m}} \theta(r),
}
using the result of the previous section. Similarly,
\eqa{
\xi 
&=& \frac{1}{2^{m}}\sum_{P\in\P}\pr{\a}(\a)\pr{\d}(\d)\sum_{\ScoR}\sum_{\begin{array}{l}\scriptstyle\cR\in X_{\aR}\st\\\scriptstyle \w(\cR) < \dw/2 \end{array}} \scalarb{E_{\c}}{E_{\c}}\\
&\leq& \frac{1}{2^{m}}.
}

Consequently,
\eqa{
\lefteqn{\sum_{P\in\P}\sum_{v\in\V_P}\sum_{\k\in\bin^m}\left|\pr{\k v}(\k,v) - \frac{1}{2^m}\pr{v}(v)\right|}\nonumber\\
&\leq&  2\left(\theta(r) + 2\sqrt{\theta(r)}\right)
}
which concludes our proof.\fin


\subsection{Bound on the conditional entropy}

We conclude the proof of privacy by using the following property from classical information theory.

\begin{property}
Let $\rv{x}$ and $\rv{y}$ be two discrete random variables taking values in the sets $\X$ and $\Y$ respectively. Let $\mu$ be a nonnegative real number. If the following inequality is satisfied:
\eq{
\sum_{x\in\X,\,y\in\Y} \left| \pr{x y}(x,y)-\frac{1}{|\X|}\pr{y}(y) \right| \leq \mu,
}
then the conditional entropy of $\rv{x}$ given $\rv{y}$ is lower-bounded by:
\eq{
H(\rv{x} | \rv{y}) \geq (1-\mu)\log_2 |\X| -\frac{1}{\ln 2} \mu.
}
\end{property}

\proof
The hypothesis implies that there exist a set of real numbers $\eta_{x,y}$ for all $x\in\X$ and $y\in\Y$ such that:
\eq{
\pr{x y}(x,y) = \frac{1}{|\X|} \pr{y}(y) (1+\eta_{x,y}),
}
($\eta_{x,y}$ is assigned the value zero if $\pr{y}(y)=0$) obeying the inequality:
\eq{
\sum_{x\in\X, y\in\Y} \frac{1}{|\X|} \pr{y}(y) | \eta_{x,y} | \leq \mu.
}

Note that for all $x$ and $y$, we have $-1\leq\eta_{x,y}\leq|\X|-1$. Now,
\eqa{
H(\rv{x} | \rv{y}) &=& -\sum_{x\in\X,\, y\in\Y \st \pr{xy}(x,y)>0} \pr{x y}(x, y) \log_2 \pc{x}{\rv{y}=y}(x) \\
	&=& \log_2 |\X| -\sum_{x\in\X,\, y\in\Y \st \eta_{x,y}>-1} \frac{1}{|\X|} \pr{y}(y)\underbrace{\log_2(1+\eta_{x,y})}_{\leq \frac{|\eta_{x,y}|}{\ln 2}} \nonumber\\
	& & - \sum_{x\in\X,\, y\in\Y \st\eta_{x,y}>-1}\frac{1}{|\X|} \pr{y}(y)\eta_{x,y} \log_2\underbrace{(1+\eta_{x,y})}_{\leq |\X|} \\
	&\geq& \log_2 |\X| -\frac{\mu}{\ln 2}-\mu\log_2|\X|,
}
which concludes the proof.\fin

The probability distribution of the private key and the view obeys the following inequality:
\eqa{
\lefteqn{\sum_{\begin{array}{l}\scriptstyle\k\in\bin^m,\\\scriptstyle v\in\V\end{array}} \left| \pr{\k v}(\k,v)-\frac{1}{2^m} \pr{v}(v) \right|}\nonumber\\
&\leq& \sum_{P\in\P} \sum_{\begin{array}{l}\scriptstyle v\in\V_P,\\\scriptstyle\k\in\bin^m\end{array}}  \left| \pr{\k v}(\k,v)-\frac{\pr{v}(v)}{2^m} \right|+\sum_{P\notin\P} \sum_{\begin{array}{l}\scriptstyle v\in\V_P,\\\scriptstyle\k\in\bin^m\end{array}}  \left| \pr{\k v}(\k,v)-\frac{ \pr{v}(v)}{2^m} \right| \\
&\leq& 2(\theta(r)+2\sqrt{\theta(r)}) + 0.
}
where we have used the fact that the key is randomly chosen by Alice with uniform probability distribution if the validation test is not passed. Applying the above property for the random variables $\rv{\k}$ and $\rv{v}$, we obtain:
\eq{
H(\rv{\k} | \rv{v}) \geq m-2\left(m+\frac{1}{\ln 2}\right)\left(\theta(r)+2\sqrt{\theta(r)}\right),
}
which concludes the proof of privacy.\fin


\smallskip
{\bf{Acknowledgement }} The author gratefully acknowledges support provided by the European TMR Network ERP-4061PL95-1412, and thanks Hans Briegel, Artur Ekert, Nicolas Gisin, Patrick Hayden, Norbert L\"utkenhaus, Dominic Mayers, Michele Mosca, Luke Rallan, Peter Shor and Vlatko Vedral for interesting discussions and helpful comments. 




\appendix
\section{Appendix: Error detection and correction}

We describe here how we can estimate the error rate in the sifted set $S$ and then correct the discrepancies between Alice's and Bob's sifted keys using the interactive error-reconciliation scheme~\cite{BS93}.

\begin{description}

\item[Estimation of the error rate] The error rate in the sifted set can be estimated by comparing a small proportion of the bits chosen randomly in the sifted key. The compared bits should be encrypted with the one-time pad method so that a potential eavesdropper learns only the positions of the errors. A probabilistic property such as the Hoeffding inequality can be used to show that the observed error rate in the sampled proportion is not considerably lower than the error rate in the remaining part of the sifted set~\cite{May98,BBBMR99,ILM00}. For asymptotic size of the sifted set, one can take arbitrarily small but positive proportion of the sifted key for this error rate estimation.

\item[Error correction] The remaining part of the sifted set that was not sampled in the previous step must be corrected. One-way linear error-correcting codes can be used for error correction. However, they are not very efficient and considerably higher number of redundant bits are required than the Shannon limit. A practical interactive correction scheme, devised by Brassard and Salvail~\cite{BS93} gets closer to this theoretical limit. A basic description of the scheme follows:

Alice and Bob group their bits into blocks of a given size, which has to be optimised as a function of the error rate. They exchange information about the parity of each block over the public channel. These parities should be encrypted using the one-time pad method. If their parities agree then they proceed to the next block.  If their parities disagree, they deduce that there was an odd number of errors in the corresponding block, and search one of them recursively by cutting the block into two subblocks and comparing the parities of the first subblock: if the parities agree then the second subblock has an odd number of errors and if they do not, then the first subblock has an odd number of errors. Again, these parities should be encrypted. This procedure is continued recursively on the subblock with an odd number of errors. As a result of the encryption of the exchanged parities, a possible eavesdropper learns only the positions of the errors~\cite{Lut96PRA,Lut99a}.

After this first step, every considered block has either an even number of errors or none. Alice and Bob then shuffle the positions of their bits and repeat the same procedure with blocks of bigger size (this size being optimised as well).  However, when an error is corrected, Alice and Bob might deduce that some blocks treated previously now have an odd number of errors.  They choose the smallest block amongst them and correct one error recursively, as before.  They proceed until every previously treated block has an even number of errors, or none.

Similar steps follow, and the interactive error correction terminates after a specified number of steps. This number is to be optimised in order to maximise the probability that no discrepancies remain and, at the same time, minimise the number of bits used for the one-time pad encryption. Readers are referred to the original paper~\cite{BS93} for precise description and treatment of this scheme.

\end{description}

\end{document}